\documentclass[aps,reprint,amssymb,prd,floatfix,superscriptaddress,nofootinbib,dvipsnames]{revtex4-1}

\usepackage[english]{babel}
\usepackage[utf8]{inputenc}
\usepackage{comment}
\usepackage[colorinlistoftodos]{todonotes}
\usepackage{amsmath}
\usepackage{amsthm}
\usepackage{amsbsy}
\usepackage{amssymb}
\usepackage{latexsym}
\usepackage{graphicx}
\usepackage{epstopdf}
\usepackage[caption=false]{subfig}
\usepackage{xcolor}
\usepackage{makecell}
\usepackage{bm}

\makeatletter
\let\@fnsymbol\@alph
\makeatother

\begin{document}

\title{Improved limits on Fierz Interference using asymmetry measurements from the UCNA experiment.}

\author{X.~Sun}                          \affiliation{W.~K.~Kellogg Radiation Laboratory, California Institute of Technology, Pasadena, California 91125, USA}

\author{E.~Adamek}                       \affiliation{Department of Physics, Indiana University, Bloomington, Indiana 47408, USA}
\author{B.~Allgeier}                     \affiliation{Department of Physics and Astronomy, University of Kentucky, Lexington, Kentucky 40506, USA}
\author{Y.~Bagdasarova}		\affiliation{Los Alamos National Laboratory, Los Alamos, New Mexico 87545, USA} \affiliation{Department of Physics and Center for Experimental Nuclear Physics and Astrophysics, University of Washington, Seattle, Washington 98195, USA} 
\author{D.~B.~Berguno}		\affiliation{Department of Physics, Virginia Tech, Blacksburg, Virginia 24061, USA} 
\author{M.~Blatnik}                      \affiliation{W.~K.~Kellogg Radiation Laboratory, California Institute of Technology, Pasadena, California 91125, USA}

\author{T.~J.~Bowles}                    \affiliation{Los Alamos National Laboratory, Los Alamos, New Mexico 87545, USA}
\author{L.~J.~Broussard}           \thanks{Currently at Oak Ridge National Laboratory, Oak Ridge, TN 37831, USA} \affiliation{Los Alamos National Laboratory, Los Alamos, New Mexico 87545, USA}
\author{M.~A.-P.~Brown}  \thanks{Currently at Humana Military, 305 N. Hurstbourne Pkwy, Louisville, KY 40222, USA}                \affiliation{Department of Physics and Astronomy, University of Kentucky, Lexington, Kentucky 40506, USA}
\author{R.~Carr}                         \affiliation{W.~K.~Kellogg Radiation Laboratory, California Institute of Technology, Pasadena, California 91125, USA}
\author{S.~Clayton}                      \affiliation{Los Alamos National Laboratory, Los Alamos, New Mexico 87545, USA}
\author{C.~Cude-Woods}                   \affiliation{Department of Physics, North Carolina State University, Raleigh, North Carolina 27695, USA}
\author{S.~Currie}                       \affiliation{Los Alamos National Laboratory, Los Alamos, New Mexico 87545, USA}
\author{E.~B.~Dees}                      \affiliation{Department of Physics, North Carolina State University, Raleigh, North Carolina 27695, USA} \affiliation{Triangle Universities Nuclear Laboratory, Durham, North Carolina 27708, USA}
\author{X.~Ding}                         \affiliation{Department of Physics, Virginia Tech, Blacksburg, Virginia 24061, USA}
\author{B.~W.~Filippone}                 \affiliation{W.~K.~Kellogg Radiation Laboratory, California Institute of Technology, Pasadena, California 91125, USA}
\author{A.~Garc\'{i}a}                   \affiliation{Department of Physics and Center for Experimental Nuclear Physics and Astrophysics, University of Washington, Seattle, Washington 98195, USA}
\author{P.~Geltenbort}                   \affiliation{Institut Laue-Langevin, 38042 Grenoble Cedex 9, France}
\author{S.~Hasan}                        \affiliation{Department of Physics and Astronomy, University of Kentucky, Lexington, Kentucky 40506, USA}
\author{K.~P.~Hickerson}                 \affiliation{W.~K.~Kellogg Radiation Laboratory, California Institute of Technology, Pasadena, California 91125, USA}
\author{J.~Hoagland}                     \affiliation{Department of Physics, North Carolina State University, Raleigh, North Carolina 27695, USA}
\author{R.~Hong}                         \affiliation{Department of Physics and Center for Experimental Nuclear Physics and Astrophysics, University of Washington, Seattle, Washington 98195, USA}
\author{A.~T.~Holley}                    \thanks{Currently at Dept. of Physics, Tennessee Tech University, Cookeville, TN, USA}\affiliation{Department of Physics, North Carolina State University, Raleigh, North Carolina 27695, USA} \affiliation{Department of Physics, Indiana University, Bloomington, Indiana 47408, USA}
\author{T.~M.~Ito}                       \affiliation{Los Alamos National Laboratory, Los Alamos, New Mexico 87545, USA}
\author{A.~Knecht}                       \thanks{Currently at Paul Scherrer Institut, 5232 Villigen PSI, Switzerland} \affiliation{Department of Physics and Center for Experimental Nuclear Physics and Astrophysics, University of Washington, Seattle, Washington 98195, USA}

\author{C.-Y.~Liu}                       \affiliation{Department of Physics, Indiana University, Bloomington, Indiana 47408, USA}
\author{J.~Liu}                          \affiliation{Department of Physics, Shanghai Jiao Tong University, Shanghai, 200240, China}
\author{M.~Makela}                       \affiliation{Los Alamos National Laboratory, Los Alamos, New Mexico 87545, USA}
\author{R.~Mammei}                  
  \affiliation{Department of Physics, University of Winnipeg, Winnipeg, MB R3B 2E9, Canada}
\author{J.~W.~Martin}                    \affiliation{W.~K.~Kellogg Radiation Laboratory, California Institute of Technology, Pasadena, California 91125, USA}
                                         \affiliation{Department of Physics, University of Winnipeg, Winnipeg, MB R3B 2E9, Canada}
\author{D.~Melconian}                    \affiliation{Cyclotron Institute, Texas A\&M University, College Station, Texas 77843, USA}
\author{M.~P.~Mendenhall}                \thanks{Currently at Physical and Life Sciences Directorate, Lawrence Livermore National Laboratory, Livermore, CA 94550, USA}
                                         \affiliation{W.~K.~Kellogg Radiation Laboratory, California Institute of Technology, Pasadena, California 91125, USA}
\author{S.~D.~Moore}                     \affiliation{Department of Physics, North Carolina State University, Raleigh, North Carolina 27695, USA}
\author{C.~L.~Morris}                    \affiliation{Los Alamos National Laboratory, Los Alamos, New Mexico 87545, USA}
\author{S.~Nepal}                        \affiliation{Department of Physics and Astronomy, University of Kentucky, Lexington, Kentucky 40506, USA}
\author{N.~Nouri}                        \thanks{Currently at Department of Pathology, Yale University School of Medicine, New Haven, Connecticut 06510, USA}\affiliation{Department of Physics and Astronomy, University of Kentucky, Lexington, Kentucky 40506, USA}
\author{R.~W.~Pattie, Jr.}
 \thanks{Currently at Department of Physics and Astronomy, East Tennessee State University, Johnson City, TN 37814}
\affiliation{Department of Physics, North Carolina State University, Raleigh, North Carolina 27695, USA}
                                         \affiliation{Triangle Universities Nuclear Laboratory, Durham, North Carolina 27708, USA}
\author{A.~P\'{e}rez Galv\'{a}n}         \thanks{Currently at Vertex Pharmaceuticals, 11010 Torreyana Rd., San Diego, CA 92121, USA}
                                         \affiliation{W.~K.~Kellogg Radiation Laboratory, California Institute of Technology, Pasadena, California 91125, USA}
\author{D.~G.~Phillips~II}               \affiliation{Department of Physics, North Carolina State University, Raleigh, North Carolina 27695, USA}
\author{R.~Picker}                       \thanks{Currently at TRIUMF, Vancouver, BC V6T 2A3 Canada}\affiliation{W.~K.~Kellogg Radiation Laboratory, California Institute of Technology, Pasadena, California 91125, USA}
\author{M.~L.~Pitt}                      \affiliation{Department of Physics, Virginia Tech, Blacksburg, Virginia 24061, USA}
\author{B.~Plaster}                      \affiliation{Department of Physics and Astronomy, University of Kentucky, Lexington, Kentucky 40506, USA}
\author{D.~J.~Salvat}                    \affiliation{Department of Physics and Center for Experimental Nuclear Physics and Astrophysics, University of Washington, Seattle, Washington 98195, USA}
\author{A.~Saunders}                     \affiliation{Los Alamos National Laboratory, Los Alamos, New Mexico 87545, USA}
\author{E.~I.~Sharapov}		\affiliation{Joint Institute for Nuclear Research, 141980, Dubna, Russia} 
\author{S.~Sjue}                         \affiliation{Los Alamos National Laboratory, Los Alamos, New Mexico 87545, USA}
\author{S.~Slutsky}                      \affiliation{W.~K.~Kellogg Radiation Laboratory, California Institute of Technology, Pasadena, California 91125, USA}
\author{W.~Sondheim}                     \affiliation{Los Alamos National Laboratory, Los Alamos, New Mexico 87545, USA}
\author{C.~Swank}                        \affiliation{W.~K.~Kellogg Radiation Laboratory, California Institute of Technology, Pasadena, California 91125, USA}
\author{E.~Tatar}                        \affiliation{Department of Physics, Idaho State University, Pocatello, Idaho 83209, USA}
\author{R.~B.~Vogelaar}                  \affiliation{Department of Physics, Virginia Tech, Blacksburg, Virginia 24061, USA}
\author{B.~VornDick}                     \affiliation{Department of Physics, North Carolina State University, Raleigh, North Carolina 27695, USA}
\author{Z.~Wang}                         \affiliation{Los Alamos National Laboratory, Los Alamos, New Mexico 87545, USA}
\author{W.~Wei}
 \affiliation{W.~K.~Kellogg Radiation Laboratory, California Institute of Technology, Pasadena, California 91125, USA}
\author{J.~W.~Wexler}                       \affiliation{Department of Physics, North Carolina State University, Raleigh, North Carolina 27695, USA}
\author{T.~Womack}                       \affiliation{Los Alamos National Laboratory, Los Alamos, New Mexico 87545, USA}
\author{C.~Wrede}                        \affiliation{Department of Physics and Center for Experimental Nuclear Physics and Astrophysics, University of Washington, Seattle, Washington 98195, USA}
                                         \affiliation{Department of Physics and Astronomy, Michigan State University, East Lansing, Michigan 48824, USA}
\author{A.~R.~Young}                     \affiliation{Department of Physics, North Carolina State University, Raleigh, North Carolina 27695, USA}
                                         \affiliation{Triangle Universities Nuclear Laboratory, Durham, North Carolina 27708, USA}
\author{B.~A.~Zeck}                      \affiliation{Department of Physics, North Carolina State University, Raleigh, North Carolina 27695, USA}

\collaboration{UCNA Collaboration}
\noaffiliation

\date{\today}

\begin{abstract}
The Ultracold Neutron Asymmetry (UCNA) experiment was designed to measure the $\beta$-decay asymmetry parameter, $A_0$, for free neutron decay. In the experiment, polarized ultracold neutrons are transported into a decay trap, and their $\beta$-decay electrons are detected with $\approx 4\pi$ acceptance into two detector packages which provide position and energy reconstruction. The experiment also has sensitivity to $b_{n}$, the Fierz interference term in the neutron $\beta$-decay rate. In this work, we determine $b_{n}$ from the energy dependence of $A_0$ using the data taken during the UCNA 2011-2013 run. In addition, we present the same type of analysis using the earlier 2010 $A$ dataset. Motivated by improved statistics and comparable systematic errors compared to the 2010 data-taking run, we present a new $b_{n}$ measurement using the weighted average of our asymmetry dataset fits, to obtain $b_{n} = 0.066 \pm 0.041_{\text{stat}} \pm 0.024_{\text{syst}}$ which corresponds to a limit of $-0.012 < b_{n} < 0.144$ at the 90\% confidence level.
\end{abstract}

\maketitle
Standard Model predictions of the electroweak sector can be tested using precision measurements of nuclear $\beta$-decay and free neutron $\beta$-decay parameters. There are many past and current experiments to measure these decay parameters such as lifetimes, angular/spin correlations, and energy spectra, to name a few \cite{Erler2004,Severijns2006,Severijns_ann_rev2011,Bhattacharya2012,Gonzalez_N-C2013, Young2014mxa,Baessler2014}. The Fierz interference term is one such decay parameter in the neutron $\beta$-decay rate (explained below) which vanishes in the Standard Model but would serve as a probe for beyond Standard Model physics in scalar and tensor couplings \cite{Erler2004, Bhattacharya2012, Gonzalez-Alonso:2018omy, Hickerson:2017fzz}.


In this publication, the UCNA collaboration presents a measurement of the Fierz interference term, $b_{n}$, using the energy-dependence of the neutron $\beta$-decay asymmetry for the 2010 \cite{Mendenhall2013}, and 2011-2013 data-taking runs \cite{Brown:2017mhw}. For completeness, the electron energy spectra from the 2011-2013 data-taking runs are also used to extract $b_{n}$ as was done for the 2010 dataset \cite{Hickerson:2017fzz}. The experiment was designed to extract the neutron $\beta$-decay asymmetry parameter, $A_{0}$, and as such was not optimized for a spectral measurement of neutron decay products. 
Thus, as shown in \cite{Hickerson:2017fzz}, the systematic uncertainty is much larger than the statistical uncertainty when $b_{n}$ is extracted from the energy dependence of the decay spectrum. In contrast, when $b_{n}$ is extracted from the energy dependence of the decay asymmetry $A(E)$ the systematic uncertainty can be much smaller than the statistical uncertainty, as shown in this analysis.

In the Standard Model for neutron $\beta$-decay the free neutron will decay with a half-life of approximately 15 minutes \cite{Greene:2016, RevModPhys.83.1173} via the decay channel $n \rightarrow p + e^- + \overline{\nu}_{e}$. The differential neutron decay rate, $d\Gamma$, contains a set of correlation coefficients which relate the various underlying weak interaction effects to the outgoing kinematics of the decay products (the proton $p$, $\beta$-decay electron $e^-$, and the electron antineutrino $\overline{\nu}_{e}$). Expressing the neutron decay rate in terms of the neutron spin,  $\vec{\sigma}_n=\vec{J}_{n}/|\vec{J}_{n}|$, and
momenta, $\vec{p}_{e},\vec{p}_{\nu}$, and total energies, $E_e,E_\nu$, of the final state particles gives \cite{Jackson1957a}:
\begin{align}
\label{eq:neutron_decay_rate}
\begin{split}
	d\Gamma = 
	\mathcal{W}(E_e)
	\Bigl[
		1 
		&+ a \, \frac{\vec{p}_e \cdot \vec{p}_\nu}{E_e E_\nu} 
		+ b_n \frac{m_e}{E_e} 
		+ A \frac{\vec{p}_e \cdot \vec{\sigma}_n}{E_e} \\
		&+ B \, \frac{\vec{p}_\nu \cdot \vec{\sigma}_n}{E_\nu}
        + \cdots ~
	\Bigr] ~ dE_e dE_\nu\,d\Omega_e d\Omega_\nu, 
\end{split}
\end{align}
where $\mathcal{W}(E_e)$ includes the total decay rate (e.g. $1/\tau_n$) and the phase space along with recoil-order, radiative and Coulomb corrections. The correlation coefficients also include recoil-order corrections. 

The $A$ coefficient is termed the beta asymmetry coefficient and was the original goal of the UCNA experiment \cite{Mendenhall2013, Mendenhall2014, Brown:2017mhw, Brown2018, plaster2019final, Liu2010, Pattie2009}. Since these $A$ measurements required timing, position, and energy reconstruction of $\beta$-decay electrons, they also allow for a spectral measurement of neutron $\beta$-decay.

The $b_n$ term in Eq. \ref{eq:neutron_decay_rate} corresponds to an energy distortion of the neutron $\beta$-decay spectrum and hence would provide a signature in any measurement of kinematic quantities. Namely, $b_{n}$ survives after integrating over all angle and energy variables for the proton and the antineutrino \cite{Gonzalez-Alonso:2018omy}:
\begin{align}
\label{eq:SM_plus_Fierz_only}
	d\Gamma_b(E_e) = \left(1 + b_n \frac{m_e}{E_e}\right)\mathcal{W}(E_e)~dE_e.
\end{align}
where $E_e$ is the electron's total energy.

In the Standard Model with the presence of only V-A (vector axial-vector) interactions, $b_{n}=0$ \cite{Gonzalez-Alonso:2018omy} (noting, again, that energy-dependent radiative and recoil order terms are absorbed in $\mathcal{W}$). Hence, $b_{n} \neq 0$ serves as a probe of beyond Standard Model scalar or tensor couplings/interactions. Furthermore, setting a limit on $b_{n}$ is important for all other correlation coefficient measurements since they would, in principle, be distorted by the presence of a $b_{n} \neq 0$ energy distortion \cite{Gonzalez-Alonso:2018omy}.

This work utilizes the data taken by the UCNA experiment in data-taking runs 2010, 2011-2012, and 2012-2013 by the UCNA collaboration.
The UCNA experiment is located at Los Alamos National Laboratory, at the Utracold Neutron facility at the Los Alamos Neutron Science Center. Previous works have extensively detailed the UCNA experiment \cite{Plaster2012, Hickerson:2017fzz, Mendenhall2013, Mendenhall2014, Brown:2017mhw, Brown2018}.
As a short overview, neutrons are produced from a tungsten spallation target \cite{Saunders2003,Morris2002, Saunders2013}, cooled to ultracold neutron (UCN) energies ($<350$~neV), polarized and transported to a main spectrometer \cite{PlasterSCS2008, ItoMWPC2007}. Within the spectrometer, UCNs undergo $\beta$-decay and the decay electrons are directed towards two detectors on either end of the spectrometer by a 1~T magnetic field, effectively giving $4\pi$ acceptance of $\beta$-decay electrons. The two detectors are hereafter denoted detectors ``1'' and ``2''.

In the UCNA apparatus, four detector count rates are measured corresponding to the two detectors and the two neutron spin directions\footnote{In the UCNA experiment, neutrons are loaded in sequence with their spins aligned or anti-aligned with the magnetic field. The ordering, in addition to background runs and depolarization runs, is grouped together into an ``octet''. More details can be found in \cite{Plaster2012}.}. 
These rates can be written as a function of decay electron total energy, $E_e$, and angle, $\theta$, between the neutron spin and electron momentum by using Eq. \ref{eq:neutron_decay_rate} and integrating over the neutrino momentum:
\begin{equation}
	\begin{aligned}
	\label{eq:rate-model}
    r^\uparrow_1(E_e) &= \tfrac{1}{2} \eta_1(E_e) 
		N^\uparrow\left ( 1 + b_n m_e/E_e + Ay(E_e)\right )
			 \, \mathcal{W}(E_e),
	\\
    r^\uparrow_2(E_e) &= \tfrac{1}{2} \eta_2(E_e) 
		N^\uparrow \left (1 + b_n m_e/E_e - Ay(E_e)\right )
           	\, \mathcal{W}(E_e),
    \\
    r^\downarrow_1(E_e) &= \tfrac{1}{2} \eta_1(E_e) 
		N^\downarrow \left ( 1 + b_n m_e/E_e - Ay(E_e)\right )
           \, \mathcal{W}(E_e),
	\\
    r^\downarrow_2(E_e) &= \tfrac{1}{2} \eta_2(E_e) 
		N^\downarrow \left ( 1 + b_n m_e/E_e + Ay(E_e)\right )
            \, \mathcal{W}(E_e),
	\end{aligned}
\end{equation}
where, for example, $r^\uparrow_2$ corresponds to the rate in detector 2 for spin $\uparrow$ (neutron polarization aligned with the imposed magnetic field), 
    \, 
    $y(E_e) \equiv \left<P\right> \beta \left<\cos \theta\right>$,
with $\left<P\right>$ the average polarization, and $\beta = v/c$ with $v$ the $\beta$-decay electron velocity, and $c$ the speed of light.
These four rates are expressed in terms of the detector efficiencies,
$\eta_{1,2}(E_e)$, and the number of stored UCN for the spin states, $N^{\uparrow,\downarrow}$.

The super-ratio \cite{Mendenhall2014, Brown2018}, hereafter denoted SR, can be defined in terms of the measured, energy-dependent, detector count rates in the 1-2 detectors for the two spin states \cite{Plaster2012, Brown:2017mhw}, $r^{\uparrow(\downarrow)}_{1(2)}(E_e)$:
\begin{align}
\label{eq:super_ratio}
    SR &= \frac{r^{\downarrow}_{1}(E_e) r^{\uparrow}_{2}(E_e)}{r^{\uparrow}_{1}(E_e) r^{\downarrow}_{2}(E_e)}
\end{align}
In this definition of the super-ratio, differences in detector efficiencies (parametrized by the $\eta_{1,2}(E_e)$) and differences in integrated counts between the spin states (parametrized by $N^{\uparrow,\downarrow}$) cancel out. Energy-dependent non-linearities are also suppressed at first-order.

The asymmetry as a function of energy can then be calculated from the super-ratio as
\begin{align}
    A_{\text{measured}}(E_e) &= \frac{1-\sqrt{SR}}{1+\sqrt{SR}} = P_n A_0 \beta \left <\cos\theta\right >
\end{align}
and the asymmetry parameter, $A_0$, can be extracted once the terms $\beta$, $\cos\theta$, and polarizations $P_n$ are known \cite{Brown:2017mhw, Brown2018}.

When $b_{n}$ is non-zero,
$A_{0}$ acquires an energy-dependent distortion \cite{Gonzalez-Alonso:2018omy}:
\begin{align}
\label{eq:asymm_b}
    A_{0,b}(E_e) &= \frac{A_{0}}{1 + b_{n}\frac{m_e}{E_e}}
\end{align}
We can therefore fit the measured asymmetry as a function of total electron energy, $E_e$, to find $b_{n}$. We note that $A_{0}, b_{n}$ are the only free parameters.

The systematic uncertainties in our analysis are quite different for the two $b_{n}$ extraction methods presented here: fitting the  energy-dependence of the asymmetry (Eq. \ref{eq:asymm_b}) vs. fitting the electron energy spectrum shape (Eq. \ref{eq:SM_plus_Fierz_only}). In the asymmetry data fit, most energy calibration systematic uncertainties are suppressed (to first order) in the super-ratio, Eq. \ref{eq:super_ratio}.
We will examine several potential sources of systematic uncertainty and their contribution to the asymmetry data. We first discuss the energy calibration uncertainty since it dominates the uncertainty in the spectral fit, and provides a suppressed but non-trivial correction and uncertainty to the asymmetry data fit. 

\begin{figure}
\includegraphics[width=\columnwidth]{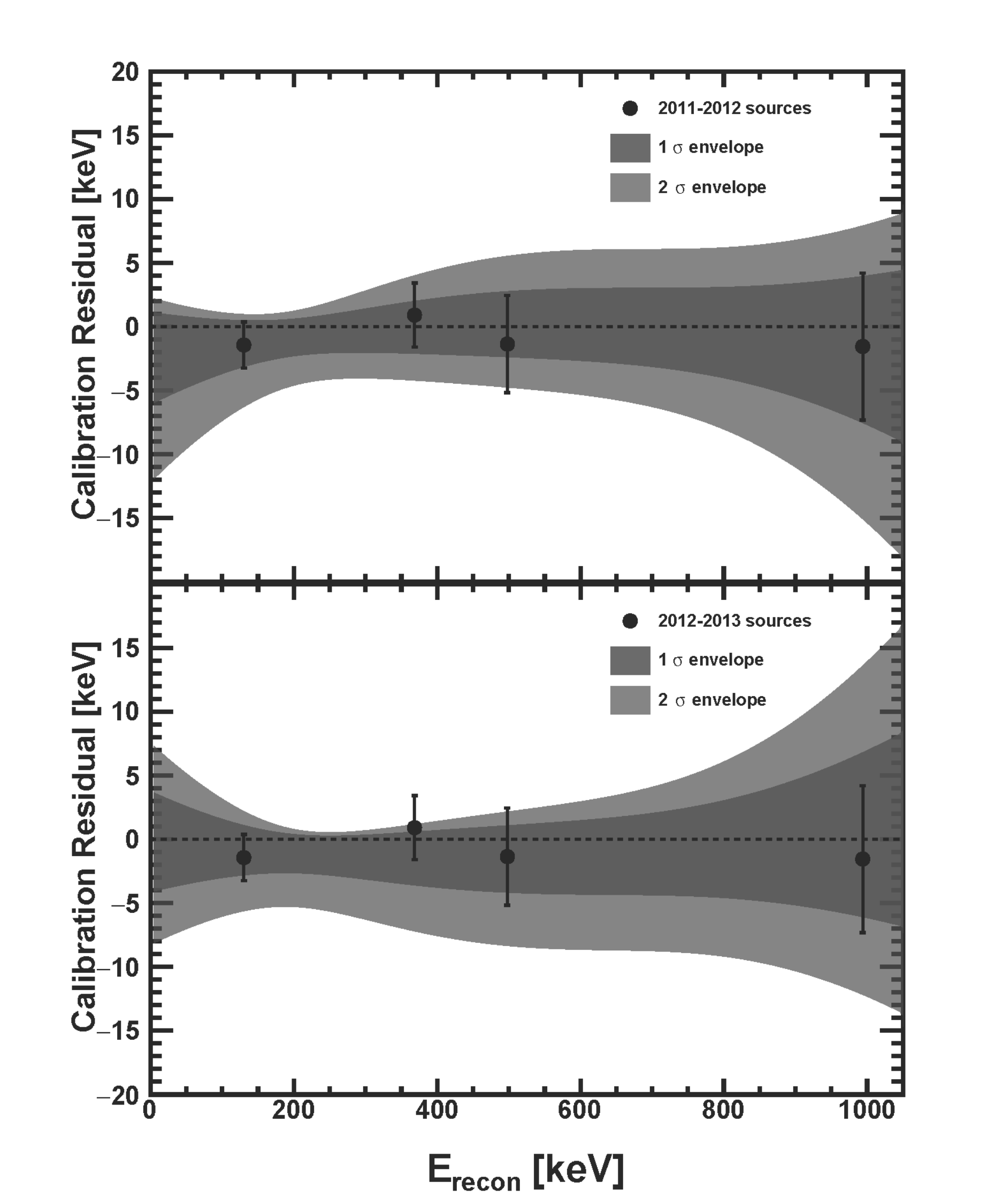}
\caption{\label{fig:error_envelopes}Error envelopes used in the analysis to generate energy calibration variations. Each error envelope is plotted with 1- and 2-$\sigma$ bands. Note that in the actual sampling of energy calibration variations, up to 3-$\sigma$ bands are used. The 2010 error envelope can be found in \cite{Hickerson:2017fzz}.}
\end{figure}

Using the method described in \cite{Brown:2017mhw, Hickerson:2017fzz}, we can construct an energy uncertainty envelope from the calibration source data-taking runs during 2011-2013. 
In this analysis, we use an asymmetric error envelope. 
The final error envelopes chosen are shown in Fig. \ref{fig:error_envelopes}.
The energy calibration for 2012-2013 has wider error bands than the 2011-2012 energy calibration. This is likely due to several factors: there were less low-energy calibration runs taken in 2012-2013 and one of the PMTs on detector 2 was missing a gain monitor for the duration of the 2012-2013 run. 
Ultimately, this may have affected the energy reconstruction.

In order to estimate the systematic uncertainty due to the energy calibration uncertainty, we examine the distribution of extracted $b_{n}$ values when applying a statistical distribution of different energy calibration variations.
As in \cite{Hickerson:2017fzz}, non-linear calibration variations, up to second-order polynomials, are sampled and accepted with relative probabilities based on whether they populate the 1-, 2-, or 3-$\sigma$ bands of the energy error envelope. These calibration variation polynomials are then checked by comparing their energy calibration residual distribution (namely, the width of the residual distribution) against the width of the measured error envelope at $131$~keV, $368$~keV, $511$~keV, $998$~keV, which corresponds to the mean energy for our four calibration source peaks of $^{137}$Ce, $^{113}$Sn, and $^{207}$Bi ($^{207}$Bi has two energy peaks for calibration).
Each energy calibration variation is then re-sampled against a theoretical $\chi^2$ distribution with the number of degrees of freedom equal to one. By forcing the distribution of variations to obey a $\chi^2$ distribution, the re-sampled distribution of energy calibration variations can be approximated as statistical.
The final energy calibration variations produced after $\chi^2$ re-sampling are used to estimate the spread in $b_{n}$ due to the energy calibration uncertainty. 

\begin{figure}
\includegraphics[width=\columnwidth]{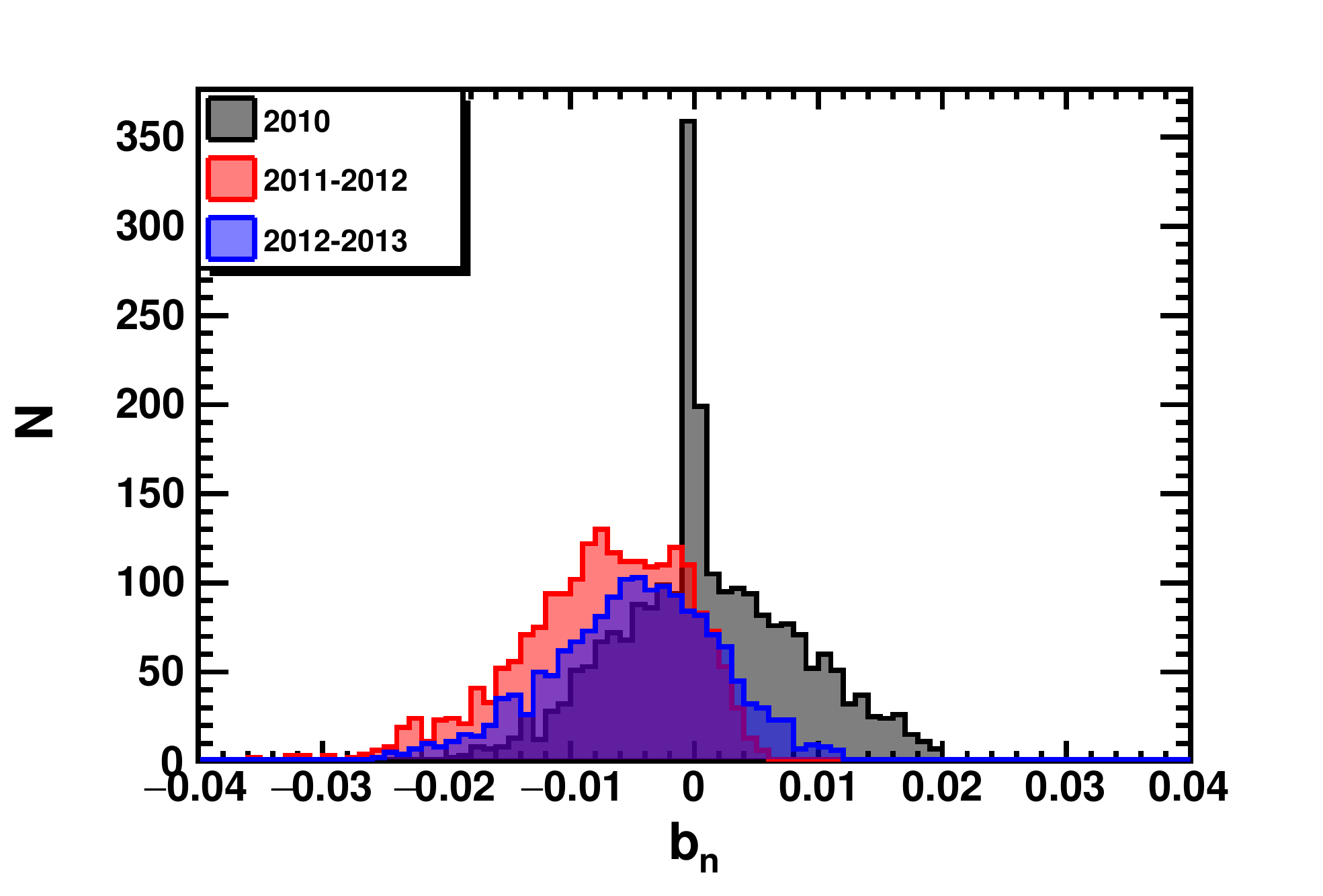}
\caption{Distribution of each year's energy calibration variations applied prior to $b_{n}$ fit from the asymmetry data. This is Monte Carlo asymmetry data with $b_{n}=0$ and central value of $A_0$ set to each dataset's extracted (Eq. \ref{eq:asymm_b}) $A_0$ value. Note the 2010 distribution is peaked at $b_{n}=0$ because the symmetric error envelope (see \cite{Hickerson:2017fzz}) allows for more energy calibration variations near the polynomial with zero variation.
}
\label{fig:twiddle_SR_bFits}
\end{figure}

Due to the asymmetric uncertainty envelopes, the energy calibration variations introduce a bias in fitted $b_{n}$ values. We estimate this by taking a $b_{n}=0$ Monte Carlo asymmetry (with statistical errors given by the data) and first converting $A_0$ into $A(E)$ by multiplying by energy-dependent terms such as $\beta = v/c$, weak magnetism, recoil, and radiative corrections.
Next, the binned energy centers are shifted according to the energy calibration variation. Finally, the new, varied $A(E)'$ is divided by the aforementioned energy dependent effects with new energy $E'$ to recover an $A_0'$. The distribution of energy calibration variation effects on the extraction of $b_{n}$ for the asymmetry data can be seen in Fig. \ref{fig:twiddle_SR_bFits}. The bias in the mean $b_{n}$ fit after applying the calibration variations is applied as a correction. We ultimately apply a bias correction of 0.0050 for 2011-2012, 0.0075 for 2012-2013 asymmetry dataset extractions. Note the 2010 dataset has a symmetric error envelope so there is no bias correction. 
Furthermore, we check the spread in $b_{n}$ bias for different Monte Carlo input $b_{n} = \pm \sigma_{b, \text{stat}}$. This small spread in bias is added as an  additional uncertainty: $\approx 0.0005$ (2011-2012), $\approx 0.0006$ (2012-2013). It is conservatively assumed to be correlated with and hence added linearly to the energy calibration variation error, and the final results are shown in the ``Energy Response'' entry in Table \ref{tab:systematics}.

In addition, various systematic effects which could influence the $b_{n}$ fit to the super-ratio asymmetry data are examined: electron backscattering and angle correction, background subtraction, detector efficiency, energy resolution, inner Bremsstrahlung, along with the aforementioned energy calibration variation (see Table \ref{tab:systematics} for a summary). Limits were set based on our GEANT4 simulations used in previous analyses \cite{Mendenhall2014, Brown2018}.

The dominant uncertainties in the asymmetry analysis result from $\beta$-decay electron scattering effects. In the UCNA detector, $\beta$-decay electrons can ``backscatter'' - scatter off materials and trigger either multiple detectors or a detector opposite its initial direction of propagation. In addition, there is an energy loss associated with the angular acceptance of $\beta$-decay electrons, denoted the $\cos\theta$ effect (more details in \cite{Plaster2012, Brown2018}). These two effects account for the dominant energy-dependent systematic uncertainty in $A_0$ and must therefore be accounted for in a $b_{n}$ extraction from the asymmetry. The systematic shift due to these effects is estimated by looking at the uncertainty in the correction applied to the asymmetry. 
For backscattering, we take a maximum +(-) 0.31\% deviation in $A_{0,b}$ at the low end of the fit region and a corresponding -(+) 0.31\% deviation at the high end, as determined in \cite{Brown:2017mhw, Brown2018}. We use a linear distortion to the asymmetry from these points to get a deviation in the slope of $A_{0,b}$, and fit to Eq. \ref{eq:asymm_b} to get a $1\sigma_b$ interval for $b_{n}$, which yields $\sigma_{b, \text{backscattering}} = 0.013$.
For energy loss due to angular acceptance, we repeat the same procedure with a maximum +(-) 0.31\% deviation at the low end of the fit region and a corresponding -(+) 0.51\% deviation at the high end, which yields $\sigma_{b, \cos\theta} = 0.017$. We note that we conservatively took the larger uncertainty in Monte Carlo corrections between 2011-2012, 2012-2013 but that the different uncertainties ultimately give $\Delta\sigma_b \approx 0.001$.

To estimate the $1 \sigma$ systematic uncertainty due to the background, we took $2 \times 10^{7}$ simulated neutron $\beta$-decay events (with spin up/down) and applied a $1 \sigma$ shift to the two detector rates based on the background model used in the analysis of $A$ \cite{Brown2018}. This shifted the counts in every bin in both detectors\footnote{While the actual background model used rates, conversion to counts was done using the live-time ratios.}, for both spin states, higher and lower by $1 \sigma$ based on Gaussian counting statistics, which apply to the background model. Fitting with Eq. \ref{eq:asymm_b} limits the uncertainty at $\sigma_{b, \text{background}} < 0.009$.

The systematic effect on $\beta$ event acceptance due to detector efficiency is estimated by simulating a variation of $\pm 20\%$ on the inefficiency of the detector, which was the same method as in \cite{Hickerson:2017fzz}. This gives $\sigma_{b, \text{efficiency}} = 0.002$. We note that the efficiency in 2011-2012, 2012-2013 is $ > 98\%$ above the low-energy cut region in this analysis, compared to $ > 90\%$ in \cite{Hickerson:2017fzz}, due to the 40~keV higher low-energy cut. We note that in \cite{Hickerson:2017fzz} the backscattering and $\cos\theta$ corrections were implicitly encompassed in the Monte Carlo detector efficiency. However, for the asymmetry data, a separate correction was applied and hence those systematic uncertainties are separated out.

The systematic uncertainty due to energy resolution was estimated by ``smearing'' each simulated event's energy by a Gaussian with its width given by the energy resolution at that event's true energy. The energy resolution for both years' dataset was $\approx 7\%$ at kinetic energy 1~MeV. A reasonable maximum variation of $\pm 10\%$ on the energy resolution gives $\sigma_{b, \text{resolution}} = 0.0002$.

The effect of detecting photons from inner Bremsstrahlung as a potential energy distortion in our spectral measurements can be estimated by noting that the UCNA detector has an efficiency of $\approx10^{-4}$ for $400$~keV gamma rays to deposit $>0$~keV energy, taken from a GEANT4 simulation. Additionally, there is a solid angle suppression factor of $\approx10^{-5}$. Finally, applying the branching ratio for detectable photons measured in \cite{Bales2016} ($\approx 3\times10^{-3}$), we conclude that for $\approx5.3\times10^{7}$ decays a negligible number would have produced a Bremsstrahlung $\gamma + \beta$ coincidence in our detector. We note that any energy distortion due to undetected photons from inner Bremsstrahlung is already incorporated in the GEANT4 simulation used in this analysis.

\begin{table}
\begin{center}
\begin{tabular}{l|c}
\hline
\hline
Type of uncertainty & Systematic uncertainty on $b_{n}$  \\
\hline
Energy Response \qquad &  $\sigma_b = 0.007$  \\
Electron Backscattering\qquad  &  $\sigma_b =  0.013$ \\
$\cos\theta$ Energy Loss\qquad & $\sigma_b = 0.017$  \\
Background Subtraction \qquad &  $\sigma_b < 0.009$ \\
Detector Efficiency \qquad &  $\sigma_b = 0.002$ \\
Energy Resolution\qquad &  $\sigma_b = 0.0002$ \\
\hline
\hline
\end{tabular}
\caption{Summary of systematic uncertainties on $b_{n}$ greater than $10^{-4}$. The energy calibration variation uncertainty is computed for different error envelopes (see Fig. \ref{fig:error_envelopes}), however the values are ultimately the same for all three datasets.}
\label{tab:systematics}
\end{center}
\end{table}

\begin{figure}[h]
\includegraphics[width=\columnwidth]{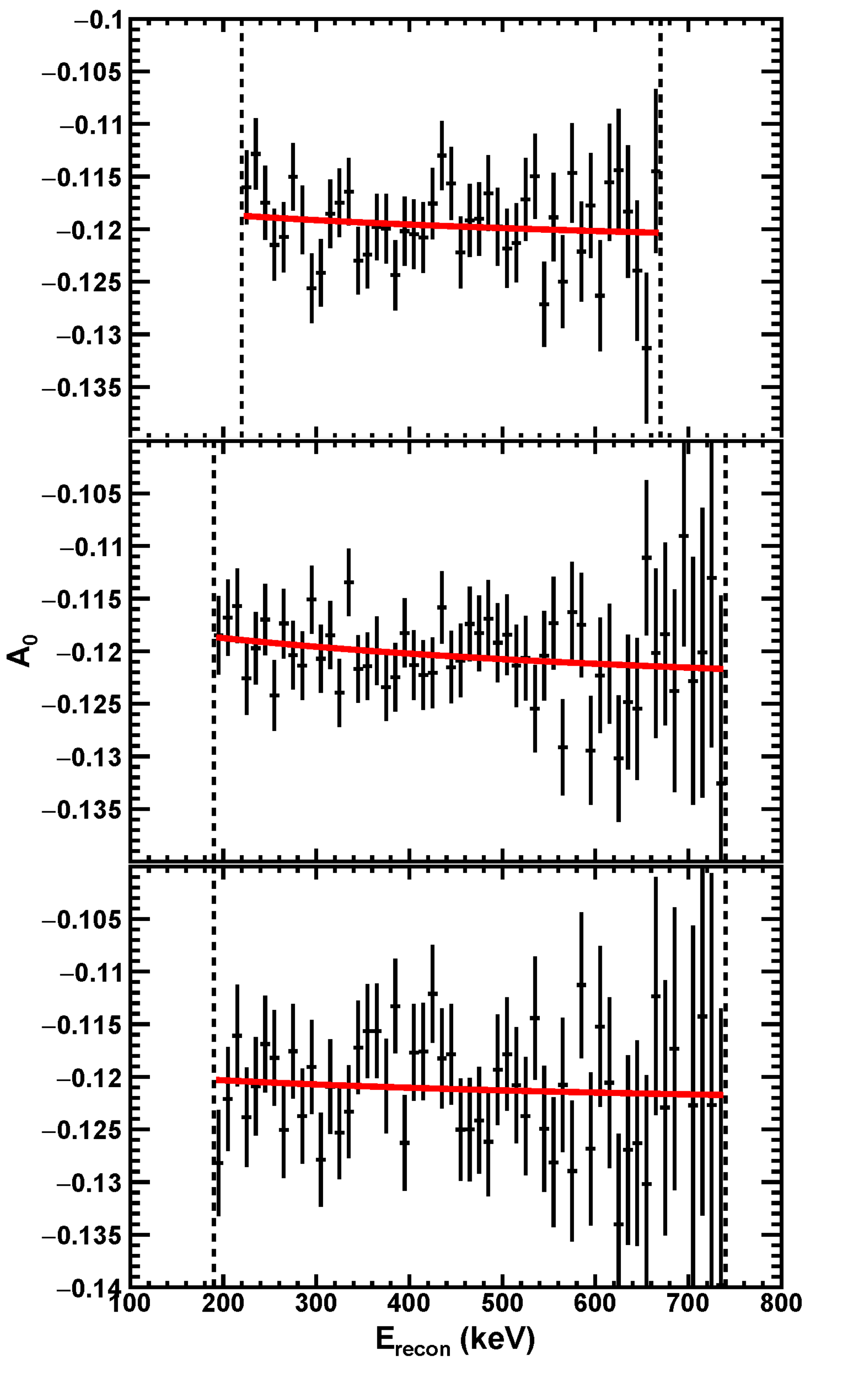}
\caption{\label{fig:asymmetry_data}Asymmetry data shown for (top) 2010, (middle) 2011-2012, and (bottom) 2012-2013 datasets. Corresponding fit function Eq. \ref{eq:asymm_b} plotted in red. The vertical dashed lines denote the energy fit region (see text). Only the data within the energy fit region is shown.}
\end{figure}

The asymmetry data published in \cite{Mendenhall2013, Brown:2017mhw} is used to extract a measurement of $b_{n}$, with all corrections being applied for the following effects: missed backscatter events, $\cos\theta$ effects, depolarization, and theoretical radiative and recoil order corrections. This data with the fit function given by Eq. \ref{eq:asymm_b} is shown in Fig. \ref{fig:asymmetry_data}. Here, the error bars are purely statistical.

The analysis described in this publication was blinded. Blinding was done on the asymmetry data by selecting an unknown $b_{\text{blind}} \in [-0.075, 0.075]$ and multiplying the asymmetry data as a function of energy by Eq. \ref{eq:asymm_b}, where $b_{n}$ is $b_{\text{blind}}$. We note this also blinds the fitted $A_0$, since it changes the average value of the asymmetry over the fitted energy range. 
The spectral data was blinded by altering the $b_{n}=0$ processed Monte Carlo used in the Eq. \ref{eq:SM_plus_Fierz_only} fit. We take an (unknown) number of events from a $b_{n}=0$ spectrum and combine them with an (unknown) number from a $b_{n}=\infty$ (i.e. Eq. \ref{eq:SM_plus_Fierz_only} with only the $\frac{m_e}{E_e}$ term) or a $b_{n}=-1$ spectrum, thus creating a final spectrum with $b_{\text{blind}}$. Note we select from a range of mixing values such that $b_{\text{blind}} \in [-0.075, 0.075]$.
All analysis cut decisions such as energy fit region, weighting of energy calibration variations, and weighted averaging procedure were decided prior to unblinding. The asymmetry datasets and the spectral datasets for 2011-2013 had different blinding factors, but each year's datasets had the same blinding factor. The 2010 asymmetry dataset was not blinded.

\begin{table*}
\renewcommand*{\arraystretch}{1.4}
\begin{center}
\begin{tabular}{l|c|c|c|c|c|c}
\hline
\hline
Measurement type & $b_{n}$ fit value & $b_{n}$ statistical error & $b_{n}$ systematic error & fit range [keV] & $\frac{\chi^2}{\text{ndf}}$ & $N_{\text{events}}$ [$10^6$] \\
\hline
2011-2012 spectrum  &  0.072 & 0.0042 & $-0.101/+0.108$  & 195-645 & 2.2 & 22\\
2012-2013 spectrum &  0.044 & 0.0079 & $-0.117/+0.174$ & 195-645 & 5.1 & 9.1\\
\hline
2011-2012 asymmetry  &  \textbf{0.087} & \textbf{0.063} & $\pm$ \textbf{0.024} & 190-740 & 0.71 & 23\\
2012-2013 asymmetry  &  \textbf{0.046} & \textbf{0.083} & $\pm$ \textbf{0.024} & 190-740 & 0.86 & 9.4\\
2010 asymmetry &  \textbf{0.052} & \textbf{0.071} & $\pm$ \textbf{0.024} & 220-670 & 0.94 & 21\\
\hline
\multicolumn{7}{c}{$\bm{\left<b_{\text{asymm}}\right> = 0.066 \pm 0.041_{\text{stat}} \pm 0.024_{\text{syst}}}$ }
\\
\hline
\hline

\end{tabular}
\caption{Summary of 1$\sigma$ fit results for 5 independent measurements of $b_{n}$. The number of events is given after all cuts are applied. The spectral extractions use the energy calibration uncertainty combined in quadrature with other systematic uncertainties estimated with the techniques in \cite{Hickerson:2017fzz}. Only the asymmetry results (bold values) are used in the weighted average.}
\label{tab:results}
\end{center}
\end{table*}

The energy range used in the $b_{n}$ extractions is the same as that which was chosen for the $A$ analysis: $220-670$~keV for 2010 \cite{Mendenhall2013, Mendenhall2014}, $190-740$~keV for 2011-2013 \cite{Brown:2017mhw, Brown2018}. This is chosen to improve our statistical power while also minimizing the various systematic effects discussed earlier. 
The dependence of the $b_{n}$ extraction on the chosen energy fit region was examined. The low energy cut, $E_{\text{low}}$, is fit for $E_{\text{low}} \pm 30$~keV for the 2011-2012, 2012-2013 asymmetry datasets. The high energy cut, $E_{\text{high}}$, is fit for $E_{\text{high}} \pm 60$~keV. The average shift in $b_{n}$ from the low energy cut variation is $\approx 0.003$, and from the high energy cut variation is $\approx 0.009$. Further systematic studies showed that the dependence of the final, weighted averaged $b_{n}$ extraction was not significantly dependent on the energy fit region as compared with the nominal energy region that was optimized in the original $A$ analysis.

Upon fitting the asymmetry datasets, we also obtain values for the asymmetry parameter $A_0$. For the 2010 dataset, we obtain $A_0 = -0.1231 \pm 0.0048$. For the 2011-2012 dataset, we obtain $A_0 = -0.1258 \pm 0.0044$. For the 2012-2013 dataset, we obtain $A_0 = -0.1236 \pm 0.0059$. In previous publications \cite{Mendenhall2013, Brown:2017mhw, plaster2019final} the reported error on $A_0$ is factor $\approx 10 \times$ smaller. This is due to the fact that under the assumption of $b=0$ in the Standard Model, there is no correlated error with a $b_{n}$ term. However, once one allows for $b_{n}$ as a free parameter, the error in the $A_0$ extraction becomes highly correlated with, and indeed dominated by the error in the $b_{n}$ extraction. We note that these results for $A_0$ agree with the previous analyses where $b=0$ was assumed. 


For comparison to the $b_{n}$ values extracted from the asymmetry data, we can also make a direct spectral distortion measurement, following the procedure of our previous publication \cite{Hickerson:2017fzz}. Namely, we can generate an energy spectrum that does not have a significant dependence on $A$ [up to $\mathcal{O}(b_n A^2$)] by forming a super-sum as the sum of the geometric means of the spin/detector pairs. Therefore any polarization-dependent systematic effects present in the asymmetry are largely suppressed in the super-sum spectrum, as shown in \cite{Hickerson:2017fzz}.
The systematic error in the spectrum is dominated by energy calibration uncertainty. We use the same procedure as was used to estimate the asymmetry systematic error i.e. generating energy calibration variations defined by the error envelopes in Fig. \ref{fig:error_envelopes}, and examining the distribution of $b_{n}$ fits afterwards. The bias induced in $b_{n}$ due to these variations is also estimated by examining several Monte Carlo simulations with various input $b_{\text{input}} \in [-0.1, 0.1]$. 
Other systematic effects are estimated using the techniques in \cite{Hickerson:2017fzz} and those uncertainties are included in the value shown in Table \ref{tab:results}.
The fit windows were chosen on the low-end to avoid the trigger function (matching the $A$ analysis) but cut out more of the higher-energy spectrum since there was low sensitivity to $b_{n}$ and increasingly larger energy resolution systematics. 
After fitting the spectral data to Eq. \ref{eq:SM_plus_Fierz_only}, we apply the bias corrections described previously: -0.024 for 2011-2012, 0.158 for 2012-2013. An additional systematic error is added for the spread in biases $\Delta b \approx 0.012$ for 2011-2012, $\Delta b \approx 0.009$ for 2012-2013. The same correlation reasoning is applied as in the $b_{n}$ bias correction for the asymmetry dataset and therefore this spread in bias is added linearly to the energy calibration systematic uncertainty.

The final unblinded fit results are shown in Table \ref{tab:results}. Based on these fit results, we construct a weighted average from the independent asymmetry measurements. Prior to unblinding, a decision was made to solely use the asymmetry fit data primarily due to the fact that the energy calibration systematic uncertainty was not improved between the the 2010 and 2011-2013 data-taking runs, even though the statistics were improved in the 2011-2013 data-taking runs. Since the systematic uncertainty dominates the spectral extraction of $b_{n}$ and since this uncertainty is highly correlated for the datasets, there is no improvement in the limits on $b_{n}$. Thus the spectral fit results are shown primarily to compare and contrast the difficulties with the energy calibration systematic errors in any direct $b_{n}$ measurement, as a natural extension to the analysis done in \cite{Hickerson:2017fzz}. The expected reduction in both efficiency- and calibration-related systematic errors for the asymmetry analysis is clearly demonstrated, and suggests there is at least one robust path forward for improved limits on the Fierz interference term in next-generation angular correlation experiments.

Taking the 3-point weighted average of the 3 asymmetry datasets, using a weighted error for the statistical error and a weighted average for the systematic error  \cite{HughesAndHase}, we obtain a final measurement of $b_{n} = 0.066 \pm 0.041_{\text{stat}} \pm 0.024_{\text{syst}}$. After combining the error in quadrature, this corresponds to a $90\%$ confidence interval of $b \in [-0.012, 0.144]$. These final results are presented for ultracold neutron $\beta$-decay data taken by the UCNA collaboration over the course of 3 years, with a sum total of 53 millions decays that pass all selection cuts. This is a factor 2 improvement on the limit set on the neutron Fierz interference term by the spectral analysis techniques in \cite{Hickerson:2017fzz}, using a $b_{n}$ extraction applied to the super-ratio construction of the beta asymmetry data of \cite{Mendenhall2013, Brown:2017mhw}, 
providing improve limits on possible beyond Standard Model tensor couplings.
In addition, since the limits set in this analysis are linked directly to possible Fierz-induced distortions to the energy-dependence of the measured asymmetry, they are complementary to limits derived from neutron lifetime and energy-averaged asymmetry \cite{Gardner2013, Wauters2013, Pattie:2013a, Pattie:2013b}.

This work is supported in part by the US Department
of Energy, Office of Nuclear Physics (DE-FG02-08ER41557, 
DE-SC0014622, 
DE-FG02-97ER41042) 
and the
National Science Foundation
(1002814,
1005233, 
1205977, 
1306997,
1307426, 
1506459, 
1615153,
1812340, and
1914133).
We gratefully acknowledge the support of the LDRD program (20110043DR), and the AOT division of the Los Alamos National
Laboratory.

\bibliography{references.bib}

\begin{thebibliography}{31}%
\makeatletter
\providecommand \@ifxundefined [1]{%
 \@ifx{#1\undefined}
}%
\providecommand \@ifnum [1]{%
 \ifnum #1\expandafter \@firstoftwo
 \else \expandafter \@secondoftwo
 \fi
}%
\providecommand \@ifx [1]{%
 \ifx #1\expandafter \@firstoftwo
 \else \expandafter \@secondoftwo
 \fi
}%
\providecommand \natexlab [1]{#1}%
\providecommand \enquote  [1]{``#1''}%
\providecommand \bibnamefont  [1]{#1}%
\providecommand \bibfnamefont [1]{#1}%
\providecommand \citenamefont [1]{#1}%
\providecommand \href@noop [0]{\@secondoftwo}%
\providecommand \href [0]{\begingroup \@sanitize@url \@href}%
\providecommand \@href[1]{\@@startlink{#1}\@@href}%
\providecommand \@@href[1]{\endgroup#1\@@endlink}%
\providecommand \@sanitize@url [0]{\catcode `\\12\catcode `\$12\catcode
  `\&12\catcode `\#12\catcode `\^12\catcode `\_12\catcode `\%12\relax}%
\providecommand \@@startlink[1]{}%
\providecommand \@@endlink[0]{}%
\providecommand \url  [0]{\begingroup\@sanitize@url \@url }%
\providecommand \@url [1]{\endgroup\@href {#1}{\urlprefix }}%
\providecommand \urlprefix  [0]{URL }%
\providecommand \Eprint [0]{\href }%
\providecommand \doibase [0]{http://dx.doi.org/}%
\providecommand \selectlanguage [0]{\@gobble}%
\providecommand \bibinfo  [0]{\@secondoftwo}%
\providecommand \bibfield  [0]{\@secondoftwo}%
\providecommand \translation [1]{[#1]}%
\providecommand \BibitemOpen [0]{}%
\providecommand \bibitemStop [0]{}%
\providecommand \bibitemNoStop [0]{.\EOS\space}%
\providecommand \EOS [0]{\spacefactor3000\relax}%
\providecommand \BibitemShut  [1]{\csname bibitem#1\endcsname}%
\let\auto@bib@innerbib\@empty
\bibitem [{\citenamefont {Erler}\ and\ \citenamefont
  {Ramsey-Musolf}(2005)}]{Erler2004}%
  \BibitemOpen
  \bibfield  {author} {\bibinfo {author} {\bibfnamefont {J.}~\bibnamefont
  {Erler}}\ and\ \bibinfo {author} {\bibfnamefont {M.~J.}\ \bibnamefont
  {Ramsey-Musolf}},\ }\href {\doibase 10.1016/j.ppnp.2004.08.001} {\bibfield
  {journal} {\bibinfo  {journal} {Prog. Part. Nucl. Phys.}\ }\textbf {\bibinfo
  {volume} {54}},\ \bibinfo {pages} {351} (\bibinfo {year} {2005})},\ \Eprint
  {http://arxiv.org/abs/hep-ph/0404291} {arXiv:hep-ph/0404291 [hep-ph]}
  \BibitemShut {NoStop}%
\bibitem [{\citenamefont {Severijns}\ \emph {et~al.}(2006)\citenamefont
  {Severijns}, \citenamefont {Beck},\ and\ \citenamefont
  {Naviliat-Cuncic}}]{Severijns2006}%
  \BibitemOpen
  \bibfield  {author} {\bibinfo {author} {\bibfnamefont {N.}~\bibnamefont
  {Severijns}}, \bibinfo {author} {\bibfnamefont {M.}~\bibnamefont {Beck}}, \
  and\ \bibinfo {author} {\bibfnamefont {O.}~\bibnamefont {Naviliat-Cuncic}},\
  }\href {\doibase 10.1103/RevModPhys.78.991} {\bibfield  {journal} {\bibinfo
  {journal} {Rev. Mod. Phys.}\ }\textbf {\bibinfo {volume} {78}},\ \bibinfo
  {pages} {991} (\bibinfo {year} {2006})}\BibitemShut {NoStop}%
\bibitem [{\citenamefont {Severijns}\ and\ \citenamefont
  {Naviliat-Cuncic}(2011)}]{Severijns_ann_rev2011}%
  \BibitemOpen
  \bibfield  {author} {\bibinfo {author} {\bibfnamefont {N.}~\bibnamefont
  {Severijns}}\ and\ \bibinfo {author} {\bibfnamefont {O.}~\bibnamefont
  {Naviliat-Cuncic}},\ }\href {\doibase 10.1146/annurev-nucl-102010-130410}
  {\bibfield  {journal} {\bibinfo  {journal} {Ann. Rev. Nuc. Part. Sci.}\
  }\textbf {\bibinfo {volume} {61}},\ \bibinfo {pages} {23} (\bibinfo {year}
  {2011})}\BibitemShut {NoStop}%
\bibitem [{\citenamefont {Bhattacharya}\ \emph {et~al.}(2012)\citenamefont
  {Bhattacharya}, \citenamefont {Cirigliano}, \citenamefont {Cohen},
  \citenamefont {Filipuzzi}, \citenamefont {Gonzalez-Alonso}, \citenamefont
  {Graesser}, \citenamefont {Gupta},\ and\ \citenamefont
  {Lin}}]{Bhattacharya2012}%
  \BibitemOpen
  \bibfield  {author} {\bibinfo {author} {\bibfnamefont {T.}~\bibnamefont
  {Bhattacharya}}, \bibinfo {author} {\bibfnamefont {V.}~\bibnamefont
  {Cirigliano}}, \bibinfo {author} {\bibfnamefont {S.~D.}\ \bibnamefont
  {Cohen}}, \bibinfo {author} {\bibfnamefont {A.}~\bibnamefont {Filipuzzi}},
  \bibinfo {author} {\bibfnamefont {M.}~\bibnamefont {Gonzalez-Alonso}},
  \bibinfo {author} {\bibfnamefont {M.~L.}\ \bibnamefont {Graesser}}, \bibinfo
  {author} {\bibfnamefont {R.}~\bibnamefont {Gupta}}, \ and\ \bibinfo {author}
  {\bibfnamefont {H.-W.}\ \bibnamefont {Lin}},\ }\href {\doibase
  10.1103/PhysRevD.85.054512} {\bibfield  {journal} {\bibinfo  {journal} {Phys.
  Rev.}\ }\textbf {\bibinfo {volume} {D85}},\ \bibinfo {pages} {054512}
  (\bibinfo {year} {2012})},\ \Eprint {http://arxiv.org/abs/1110.6448}
  {arXiv:1110.6448 [hep-ph]} \BibitemShut {NoStop}%
\bibitem [{\citenamefont {Naviliat-Cuncic}\ and\ \citenamefont
  {González-Alonso}(2013)}]{Gonzalez_N-C2013}%
  \BibitemOpen
  \bibfield  {author} {\bibinfo {author} {\bibfnamefont {O.}~\bibnamefont
  {Naviliat-Cuncic}}\ and\ \bibinfo {author} {\bibfnamefont {M.}~\bibnamefont
  {González-Alonso}},\ }\href {\doibase 10.1002/andp.201300072} {\bibfield
  {journal} {\bibinfo  {journal} {Annalen Phys.}\ }\textbf {\bibinfo {volume}
  {525}},\ \bibinfo {pages} {600} (\bibinfo {year} {2013})},\ \Eprint
  {http://arxiv.org/abs/1304.1759} {arXiv:1304.1759 [hep-ph]} \BibitemShut
  {NoStop}%
\bibitem [{\citenamefont {Young}\ \emph {et~al.}(2014)\citenamefont {Young}
  \emph {et~al.}}]{Young2014mxa}%
  \BibitemOpen
  \bibfield  {author} {\bibinfo {author} {\bibfnamefont {A.~R.}\ \bibnamefont
  {Young}} \emph {et~al.},\ }\href {\doibase 10.1088/0954-3899/41/11/114007}
  {\bibfield  {journal} {\bibinfo  {journal} {J. Phys.}\ }\textbf {\bibinfo
  {volume} {G41}},\ \bibinfo {pages} {114007} (\bibinfo {year}
  {2014})}\BibitemShut {NoStop}%
\bibitem [{\citenamefont {Baeßler}\ \emph {et~al.}(2014)\citenamefont
  {Baeßler}, \citenamefont {Bowman}, \citenamefont {Penttilä},\ and\
  \citenamefont {Počanić}}]{Baessler2014}%
  \BibitemOpen
  \bibfield  {author} {\bibinfo {author} {\bibfnamefont {S.}~\bibnamefont
  {Baeßler}}, \bibinfo {author} {\bibfnamefont {J.~D.}\ \bibnamefont
  {Bowman}}, \bibinfo {author} {\bibfnamefont {S.}~\bibnamefont {Penttilä}}, \
  and\ \bibinfo {author} {\bibfnamefont {D.}~\bibnamefont {Počanić}},\ }\href
  {http://stacks.iop.org/0954-3899/41/i=11/a=114003} {\bibfield  {journal}
  {\bibinfo  {journal} {Journal of Physics G: Nuclear and Particle Physics}\
  }\textbf {\bibinfo {volume} {41}},\ \bibinfo {pages} {114003} (\bibinfo
  {year} {2014})}\BibitemShut {NoStop}%
\bibitem [{\citenamefont {Gonzalez-Alonso}\ \emph {et~al.}(2019)\citenamefont
  {Gonzalez-Alonso}, \citenamefont {Naviliat-Cuncic},\ and\ \citenamefont
  {Severijns}}]{Gonzalez-Alonso:2018omy}%
  \BibitemOpen
  \bibfield  {author} {\bibinfo {author} {\bibfnamefont {M.}~\bibnamefont
  {Gonzalez-Alonso}}, \bibinfo {author} {\bibfnamefont {O.}~\bibnamefont
  {Naviliat-Cuncic}}, \ and\ \bibinfo {author} {\bibfnamefont {N.}~\bibnamefont
  {Severijns}},\ }\href {\doibase 10.1016/j.ppnp.2018.08.002} {\bibfield
  {journal} {\bibinfo  {journal} {Prog. Part. Nucl. Phys.}\ }\textbf {\bibinfo
  {volume} {104}},\ \bibinfo {pages} {165} (\bibinfo {year} {2019})},\ \Eprint
  {http://arxiv.org/abs/1803.08732} {arXiv:1803.08732 [hep-ph]} \BibitemShut
  {NoStop}%
\bibitem [{\citenamefont {Hickerson}\ \emph {et~al.}(2017)\citenamefont
  {Hickerson} \emph {et~al.}}]{Hickerson:2017fzz}%
  \BibitemOpen
  \bibfield  {author} {\bibinfo {author} {\bibfnamefont {K.~P.}\ \bibnamefont
  {Hickerson}} \emph {et~al.},\ }\href {\doibase 10.1103/PhysRevC.96.042501,
  10.1103/PhysRevC.96.059901} {\bibfield  {journal} {\bibinfo  {journal} {Phys.
  Rev.}\ }\textbf {\bibinfo {volume} {C96}},\ \bibinfo {pages} {042501}
  (\bibinfo {year} {2017})},\ \bibinfo {note} {[Addendum: Phys.
  Rev.C96,no.5,059901(2017)]},\ \Eprint {http://arxiv.org/abs/1707.00776}
  {arXiv:1707.00776 [nucl-ex]} \BibitemShut {NoStop}%
\bibitem [{\citenamefont {Mendenhall}\ \emph {et~al.}(2013)\citenamefont
  {Mendenhall} \emph {et~al.}}]{Mendenhall2013}%
  \BibitemOpen
  \bibfield  {author} {\bibinfo {author} {\bibfnamefont {M.~P.}\ \bibnamefont
  {Mendenhall}} \emph {et~al.} (\bibinfo {collaboration} {UCNA
  Collaboration}),\ }\href {\doibase 10.1103/PhysRevC.87.032501} {\bibfield
  {journal} {\bibinfo  {journal} {Phys. Rev. C}\ }\textbf {\bibinfo {volume}
  {87}},\ \bibinfo {pages} {032501} (\bibinfo {year} {2013})}\BibitemShut
  {NoStop}%
\bibitem [{\citenamefont {Brown}\ \emph {et~al.}(2018)\citenamefont {Brown},
  \citenamefont {Dees} \emph {et~al.}}]{Brown:2017mhw}%
  \BibitemOpen
  \bibfield  {author} {\bibinfo {author} {\bibfnamefont {M.~A.-P.}\
  \bibnamefont {Brown}}, \bibinfo {author} {\bibfnamefont {E.~B.}\ \bibnamefont
  {Dees}},  \emph {et~al.} (\bibinfo {collaboration} {UCNA Collaboration}),\
  }\href {\doibase 10.1103/PhysRevC.97.035505} {\bibfield  {journal} {\bibinfo
  {journal} {Phys. Rev. C}\ }\textbf {\bibinfo {volume} {97}},\ \bibinfo
  {pages} {035505} (\bibinfo {year} {2018})}\BibitemShut {NoStop}%
\bibitem [{\citenamefont {Greene}\ and\ \citenamefont
  {Geltenbort}(2016)}]{Greene:2016}%
  \BibitemOpen
  \bibfield  {author} {\bibinfo {author} {\bibfnamefont {G.~L.}\ \bibnamefont
  {Greene}}\ and\ \bibinfo {author} {\bibfnamefont {P.}~\bibnamefont
  {Geltenbort}},\ }\href {\doibase 10.1038/scientificamerican0416-36}
  {\bibfield  {journal} {\bibinfo  {journal} {Sci. Am.}\ }\textbf {\bibinfo
  {volume} {314}},\ \bibinfo {pages} {37} (\bibinfo {year} {2016})}\BibitemShut
  {NoStop}%
\bibitem [{\citenamefont {Wietfeldt}\ and\ \citenamefont
  {Greene}(2011)}]{RevModPhys.83.1173}%
  \BibitemOpen
  \bibfield  {author} {\bibinfo {author} {\bibfnamefont {F.~E.}\ \bibnamefont
  {Wietfeldt}}\ and\ \bibinfo {author} {\bibfnamefont {G.~L.}\ \bibnamefont
  {Greene}},\ }\href {\doibase 10.1103/RevModPhys.83.1173} {\bibfield
  {journal} {\bibinfo  {journal} {Rev. Mod. Phys.}\ }\textbf {\bibinfo {volume}
  {83}},\ \bibinfo {pages} {1173} (\bibinfo {year} {2011})}\BibitemShut
  {NoStop}%
\bibitem [{\citenamefont {Jackson}\ \emph {et~al.}(1957)\citenamefont
  {Jackson}, \citenamefont {Treiman},\ and\ \citenamefont
  {Wyld~Jr}}]{Jackson1957a}%
  \BibitemOpen
  \bibfield  {author} {\bibinfo {author} {\bibfnamefont {J.~D.}\ \bibnamefont
  {Jackson}}, \bibinfo {author} {\bibfnamefont {S.~B.}\ \bibnamefont
  {Treiman}}, \ and\ \bibinfo {author} {\bibfnamefont {H.~W.}\ \bibnamefont
  {Wyld~Jr}},\ }\href@noop {} {\bibfield  {journal} {\bibinfo  {journal} {Phys.
  Rev.}\ }\textbf {\bibinfo {volume} {106}},\ \bibinfo {pages} {517} (\bibinfo
  {year} {1957})}\BibitemShut {NoStop}%
\bibitem [{\citenamefont {Mendenhall}(2014)}]{Mendenhall2014}%
  \BibitemOpen
  \bibfield  {author} {\bibinfo {author} {\bibfnamefont {M.~P.}\ \bibnamefont
  {Mendenhall}},\ }\emph {\bibinfo {title} {Measurement of the neutron beta
  decay asymmetry using ultracold neutrons.}},\ \href@noop {} {Ph.D. thesis},\
  \bibinfo  {school} {California Institute of Technology} (\bibinfo {year}
  {2014})\BibitemShut {NoStop}%
\bibitem [{\citenamefont {Brown}(2018)}]{Brown2018}%
  \BibitemOpen
  \bibfield  {author} {\bibinfo {author} {\bibfnamefont {M.~A.-P.}\
  \bibnamefont {Brown}},\ }\emph {\bibinfo {title} {Determination of the
  Neutron Beta-Decay Asymmetry Parameter A Using Polarized Ultracold
  Neutrons}},\ \href@noop {} {Ph.D. thesis},\ \bibinfo  {school} {University of
  Kentucky} (\bibinfo {year} {2018})\BibitemShut {NoStop}%
\bibitem [{\citenamefont {Plaster}\ \emph {et~al.}(2019)\citenamefont {Plaster}
  \emph {et~al.}}]{plaster2019final}%
  \BibitemOpen
  \bibfield  {author} {\bibinfo {author} {\bibfnamefont {B.}~\bibnamefont
  {Plaster}} \emph {et~al.},\ }\href@noop {} {\  (\bibinfo {year} {2019})},\
  \Eprint {http://arxiv.org/abs/1904.05432} {arXiv:1904.05432 [nucl-ex]}
  \BibitemShut {NoStop}%
\bibitem [{\citenamefont {Liu}\ \emph {et~al.}(2010)\citenamefont {Liu},
  \citenamefont {Mendenhall}, \citenamefont {Holley} \emph {et~al.}}]{Liu2010}%
  \BibitemOpen
  \bibfield  {author} {\bibinfo {author} {\bibfnamefont {J.}~\bibnamefont
  {Liu}}, \bibinfo {author} {\bibfnamefont {M.~P.}\ \bibnamefont {Mendenhall}},
  \bibinfo {author} {\bibfnamefont {A.~T.}\ \bibnamefont {Holley}},  \emph
  {et~al.} (\bibinfo {collaboration} {UCNA Collaboration}),\ }\href {\doibase
  10.1103/PhysRevLett.105.181803} {\bibfield  {journal} {\bibinfo  {journal}
  {Phys. Rev. Lett.}\ }\textbf {\bibinfo {volume} {105}},\ \bibinfo {pages}
  {181803} (\bibinfo {year} {2010})}\BibitemShut {NoStop}%
\bibitem [{\citenamefont {Pattie}\ \emph {et~al.}(2009)\citenamefont {Pattie}
  \emph {et~al.}}]{Pattie2009}%
  \BibitemOpen
  \bibfield  {author} {\bibinfo {author} {\bibfnamefont {R.~W.}\ \bibnamefont
  {Pattie}} \emph {et~al.} (\bibinfo {collaboration} {UCNA Collaboration}),\
  }\href {\doibase 10.1103/PhysRevLett.102.012301} {\bibfield  {journal}
  {\bibinfo  {journal} {Phys. Rev. Lett.}\ }\textbf {\bibinfo {volume} {102}},\
  \bibinfo {pages} {012301} (\bibinfo {year} {2009})}\BibitemShut {NoStop}%
\bibitem [{\citenamefont {Plaster}\ \emph {et~al.}(2012)\citenamefont {Plaster}
  \emph {et~al.}}]{Plaster2012}%
  \BibitemOpen
  \bibfield  {author} {\bibinfo {author} {\bibfnamefont {B.}~\bibnamefont
  {Plaster}} \emph {et~al.} (\bibinfo {collaboration} {UCNA Collaboration}),\
  }\href {\doibase 10.1103/PhysRevC.86.055501} {\bibfield  {journal} {\bibinfo
  {journal} {Phys. Rev. C}\ }\textbf {\bibinfo {volume} {86}},\ \bibinfo
  {pages} {055501} (\bibinfo {year} {2012})}\BibitemShut {NoStop}%
\bibitem [{\citenamefont {Saunders}\ \emph {et~al.}(2004)\citenamefont
  {Saunders} \emph {et~al.}}]{Saunders2003}%
  \BibitemOpen
  \bibfield  {author} {\bibinfo {author} {\bibfnamefont {A.}~\bibnamefont
  {Saunders}} \emph {et~al.},\ }\href {\doibase 10.1016/j.physletb.2004.04.048}
  {\bibfield  {journal} {\bibinfo  {journal} {Phys. Lett. B}\ }\textbf
  {\bibinfo {volume} {593}},\ \bibinfo {pages} {55} (\bibinfo {year}
  {2004})}\BibitemShut {NoStop}%
\bibitem [{\citenamefont {Morris}\ \emph {et~al.}(2002)\citenamefont {Morris}
  \emph {et~al.}}]{Morris2002}%
  \BibitemOpen
  \bibfield  {author} {\bibinfo {author} {\bibfnamefont {C.}~\bibnamefont
  {Morris}} \emph {et~al.},\ }\href {\doibase 10.1103/PhysRevLett.89.272501}
  {\bibfield  {journal} {\bibinfo  {journal} {Phys. Rev. Lett.}\ }\textbf
  {\bibinfo {volume} {89}},\ \bibinfo {pages} {272501} (\bibinfo {year}
  {2002})}\BibitemShut {NoStop}%
\bibitem [{\citenamefont {Saunders}\ \emph {et~al.}(2013)\citenamefont
  {Saunders} \emph {et~al.}}]{Saunders2013}%
  \BibitemOpen
  \bibfield  {author} {\bibinfo {author} {\bibfnamefont {A.}~\bibnamefont
  {Saunders}} \emph {et~al.},\ }\href {\doibase 10.1063/1.4770063} {\bibfield
  {journal} {\bibinfo  {journal} {Review of Scientific Instruments}\ }\textbf
  {\bibinfo {volume} {84}},\ \bibinfo {pages} {013304} (\bibinfo {year}
  {2013})},\ \Eprint {http://arxiv.org/abs/https://doi.org/10.1063/1.4770063}
  {https://doi.org/10.1063/1.4770063} \BibitemShut {NoStop}%
\bibitem [{\citenamefont {Plaster}\ \emph {et~al.}(2008)\citenamefont {Plaster}
  \emph {et~al.}}]{PlasterSCS2008}%
  \BibitemOpen
  \bibfield  {author} {\bibinfo {author} {\bibfnamefont {B.}~\bibnamefont
  {Plaster}} \emph {et~al.},\ }\href {\doibase 10.1016/j.nima.2008.07.143}
  {\bibfield  {journal} {\bibinfo  {journal} {Nucl. Instrum. Meth.}\ }\textbf
  {\bibinfo {volume} {A595}},\ \bibinfo {pages} {587} (\bibinfo {year}
  {2008})},\ \Eprint {http://arxiv.org/abs/0806.2097} {arXiv:0806.2097
  [nucl-ex]} \BibitemShut {NoStop}%
\bibitem [{\citenamefont {Ito}\ \emph {et~al.}(2007)\citenamefont {Ito} \emph
  {et~al.}}]{ItoMWPC2007}%
  \BibitemOpen
  \bibfield  {author} {\bibinfo {author} {\bibfnamefont {T.~M.}\ \bibnamefont
  {Ito}} \emph {et~al.},\ }\href {\doibase 10.1016/j.nima.2006.11.026}
  {\bibfield  {journal} {\bibinfo  {journal} {Nucl. Instrum. Meth.}\ }\textbf
  {\bibinfo {volume} {A571}},\ \bibinfo {pages} {676} (\bibinfo {year}
  {2007})},\ \Eprint {http://arxiv.org/abs/physics/0702085}
  {arXiv:physics/0702085 [PHYSICS]} \BibitemShut {NoStop}%
\bibitem [{\citenamefont {Bales}\ \emph {et~al.}(2016)\citenamefont {Bales}
  \emph {et~al.}}]{Bales2016}%
  \BibitemOpen
  \bibfield  {author} {\bibinfo {author} {\bibfnamefont {M.~J.}\ \bibnamefont
  {Bales}} \emph {et~al.} (\bibinfo {collaboration} {RDK II Collaboration}),\
  }\href {\doibase 10.1103/PhysRevLett.116.242501} {\bibfield  {journal}
  {\bibinfo  {journal} {Phys. Rev. Lett.}\ }\textbf {\bibinfo {volume} {116}},\
  \bibinfo {pages} {242501} (\bibinfo {year} {2016})}\BibitemShut {NoStop}%
\bibitem [{\citenamefont {Hughes}\ and\ \citenamefont
  {Hase}(2010)}]{HughesAndHase}%
  \BibitemOpen
  \bibfield  {author} {\bibinfo {author} {\bibfnamefont {G.~I.}\ \bibnamefont
  {Hughes}}\ and\ \bibinfo {author} {\bibfnamefont {T.~P.}\ \bibnamefont
  {Hase}},\ }\href@noop {} {\emph {\bibinfo {title} {Measurements and their
  Uncertainties}}}\ (\bibinfo  {publisher} {Oxford University Press},\ \bibinfo
  {year} {2010})\BibitemShut {NoStop}%
\bibitem [{\citenamefont {Gardner}\ and\ \citenamefont
  {Plaster}(2013)}]{Gardner2013}%
  \BibitemOpen
  \bibfield  {author} {\bibinfo {author} {\bibfnamefont {S.}~\bibnamefont
  {Gardner}}\ and\ \bibinfo {author} {\bibfnamefont {B.}~\bibnamefont
  {Plaster}},\ }\href {\doibase 10.1103/PhysRevC.87.065504} {\bibfield
  {journal} {\bibinfo  {journal} {Phys. Rev. C}\ }\textbf {\bibinfo {volume}
  {87}},\ \bibinfo {pages} {065504} (\bibinfo {year} {2013})}\BibitemShut
  {NoStop}%
\bibitem [{\citenamefont {Wauters}\ \emph {et~al.}(2014)\citenamefont
  {Wauters}, \citenamefont {Garc\'ia},\ and\ \citenamefont
  {Hong}}]{Wauters2013}%
  \BibitemOpen
  \bibfield  {author} {\bibinfo {author} {\bibfnamefont {F.}~\bibnamefont
  {Wauters}}, \bibinfo {author} {\bibfnamefont {A.}~\bibnamefont {Garc\'ia}}, \
  and\ \bibinfo {author} {\bibfnamefont {R.}~\bibnamefont {Hong}},\ }\href
  {\doibase 10.1103/PhysRevC.89.025501} {\bibfield  {journal} {\bibinfo
  {journal} {Phys. Rev. C}\ }\textbf {\bibinfo {volume} {89}},\ \bibinfo
  {pages} {025501} (\bibinfo {year} {2014})}\BibitemShut {NoStop}%
\bibitem [{\citenamefont {Pattie~Jr.}\ \emph {et~al.}(2013)\citenamefont
  {Pattie~Jr.}, \citenamefont {Hickerson},\ and\ \citenamefont
  {Young}}]{Pattie:2013a}%
  \BibitemOpen
  \bibfield  {author} {\bibinfo {author} {\bibfnamefont {R.}~\bibnamefont
  {Pattie~Jr.}}, \bibinfo {author} {\bibfnamefont {K.}~\bibnamefont
  {Hickerson}}, \ and\ \bibinfo {author} {\bibfnamefont {A.}~\bibnamefont
  {Young}},\ }\href {\doibase 10.1103/PhysRevC.88.048501} {\bibfield  {journal}
  {\bibinfo  {journal} {Phys. Rev. C}\ }\textbf {\bibinfo {volume} {88}},\
  \bibinfo {pages} {048501} (\bibinfo {year} {2013})}\BibitemShut {NoStop}%
\bibitem [{\citenamefont {Pattie}\ \emph {et~al.}(2015)\citenamefont {Pattie},
  \citenamefont {Hickerson},\ and\ \citenamefont {Young}}]{Pattie:2013b}%
  \BibitemOpen
  \bibfield  {author} {\bibinfo {author} {\bibfnamefont {R.~W.}\ \bibnamefont
  {Pattie}}, \bibinfo {author} {\bibfnamefont {K.~P.}\ \bibnamefont
  {Hickerson}}, \ and\ \bibinfo {author} {\bibfnamefont {A.~R.}\ \bibnamefont
  {Young}},\ }\href {\doibase 10.1103/PhysRevC.92.069902} {\bibfield  {journal}
  {\bibinfo  {journal} {Phys. Rev. C}\ }\textbf {\bibinfo {volume} {92}},\
  \bibinfo {pages} {069902} (\bibinfo {year} {2015})}\BibitemShut {NoStop}%
\end{thebibliography}%
\end {document}